# Thermal Analysis of 3D GPU-Memory Architectures with Boron Nitride Interposer

Eric Han Wang[1], Weijia Yan[2*], Ruihong Huang[3*]

[1] *College Station High School, 4002 Victoria Ave, College Station, TX, U.S.*

[2]* *Department of Mechanical Engineering, Texas A&M University, 3123 TAMU, College Station, TX, U.S.*

[3]* *Department of Computer Science & Engineering, Texas A&M University, 3123 TAMU, College Station, TX, U.S.*

[2]*Corresponding Author Email: weijia_yan@tamu.edu

[3]*Corresponding Author Email: huangrh@tamu.edu

## Abstract

As artificial intelligence (AI) chips become more powerful, the thermal management capabilities of conventional silicon (Si) substrates become insufficient for 3D-stacked designs. This work integrates electrically insulative and thermally conductive hexagonal boron nitride (h-BN) interposers into AI chips for effective thermal management. Using COMSOL Multiphysics, the effects of High-Bandwidth Memory (HBM) distributions and thermal interface material configurations on heat dissipation and hotspot mitigation were studied. A 20 °C reduction in hot spots was achieved using h-BN interposers compared to Si interposers. Such an improvement could reduce AI chips' power leakage by 22% and significantly enhance their thermal performance.

## 1. Introduction

The rapid growth of artificial intelligence (AI) workloads has accelerated the development of high-density integrated chips that combine logic, memory, and accelerators into compact architectures[1]. Among these advances, 3D-stacked AI chips are particularly critical, as they embed more transistors and memory into a reduced footprint, delivering higher bandwidth, lower latency, and superior energy efficiency compared to conventional 2D designs[2, 3]. Within such architectures, vertically stacking the graphics processing unit (GPU) over high-bandwidth memory (HBM) is a strategic approach to eliminate memory bottlenecks, minimize data movement and latency, and significantly reduce power consumption[4]. However, 3D AI chip architectures experience significant heat accumulation, with average heat fluxes around 300 W/cm² and localized hotspots reaching 500-1000 W/cm².[5] Such extreme thermal loads pose major challenges to performance, stability, and long-term reliability. Notably, the presence of thermal hotspots significantly increases the risks of electromigration, die cracking, delamination, melting, creep, and even ignition of the packaging materials [6]. Elevated temperatures within chip stacks can also exacerbate leakage currents, compromising the accuracy and consistency of AI workloads.

Therefore, thermal management has become a critical design concern in next-generation AI hardware, particularly in 3D stacked AI chips [6, 7].

Several approaches are being explored to address the heat dissipation challenge in 3D AI chips, ranging from advanced thermal interface materials [8, 9] to embedded cooling strategies [10, 11]. Among these, the development of high-performance interposers is widely regarded as a particularly promising solution, as they can enhance heat spreading and improve thermal reliability compared to conventional approaches. Different types of interposers have been specifically designed to improve heat dissipation in 3D Integrated Circuits (ICs) and AI chip stacks[6], serving as a thermal management layer. Particularly, electrically insulating and thermally conductive materials are extremely attractive to serve as thermal management interposers because they eliminate the potential risk of electrical shorting. For example, an electrically-insulating and thermally-conductive diamond-on-chip-on-glass interposer was bonded to the silicon chips for heat dissipation, but the heat dissipation performance is dependent on the bonding strength and also subject to potential debonding at increasing temperatures [12]. Varziri et al. also investigated the thermal transport of an electrically insulating, thermally conductive aluminum nitride (AlN) film for a potential interposer; however, it has not been integrated into the chip, and no device performance has been reported [13]. Hexagonal boron nitride (h-BN) is an electrically insulating material with exceptionally high in-plane thermal conductivity 751 W/m·K[14], making it highly effective for spreading heat laterally and mitigating hotspots in 3D chip architectures. Its 2D layered structure also provides chemical stability and compatibility with advanced packaging, positioning h-BN as a promising interposer material for thermal management in next-generation AI hardware. Recently, Nazim et al. studied the chemical vapor deposition (CVD) growth of h-BN on the Pt/Cu/Ti coil and examined its electrothermal behavior [15], indicating significant potential for 3D chip thermal management. However, no attempt has been made to integrate h-BN interposers into 3D stacking AI chips, and there is a lack of fundamental understanding of the device performance for h-BN-integrated 3D AI chips.

In this study, a novel 3D AI chip consisting of stacked GPU and HBM modules, utilizing an h-BN interposer, was investigated. The h-BN interposer was used to replace traditional Silicon interposers, which face limitations in thermal conductivity and leakage under 3D AI chip conditions[16, 17]. The in-plane thermal conductivity of h-BN is approximately five times higher than that of silicon (Si), indicating its significant potential in thermal management. The key thermophysical properties of h-BN and Si used in the study are summarized in Table 1[14, 18-21].

Table 1. Thermophysical properties of h-BN and Si

| Property | h-BN | Si |
| --- | --- | --- |
| Thermal Conductivity (in-plane) | 751 W/m·K | 130–150 W/m·K |
| Thermal Conductivity (through-plane) | 2–20 W/m·K | 130–150 W/m·K |
| Coefficient of Thermal Expansion (CTE) | $1\text{–}4 \times 10^{-6}$ /K | $\sim 2.6 \times 10^{-6}$ /K |
| Specific Heat Capacity (Cp) | ~ 0.8 J/g·K | ~ 0.7 J/g·K |

COMSOL Multiphysics solvers were used to evaluate h-BN performance in interposer layers of 3D AI chips. An interposer is the dense wiring "middle layer" that routes high-speed signals and power/ground between a GPU and nearby HBM stacks. It is expected to provide short, impedance-controlled interconnects and a flat, rigid platform that aids thermal and mechanical integration. Specifically, the effects of AI chip configuration on thermal transport performance of the 3D AI chips were evaluated under typical AI chip operating conditions. The usage of h-BN is expected to lower junction temperatures in the AI chip, enhance temperature uniformity, and boost device durability.

## 2. Heat transfer model for the AI chip

Heat transfer in 3D AI chips follows the 3D steady-state heat conduction equation with internal heat generation as follows[22, 23]:

$$k\left(\frac{\partial^2 T}{\partial x^2} + \frac{\partial^2 T}{\partial y^2} + \frac{\partial^2 T}{\partial z^2}\right) + \dot{q}_g = 0 \qquad (1)$$

Where k is the thermal conductivity [W/m·K]; T is the temperature field [K]; $\dot{q}_g$ is the volumetric heat generation [W/m³], which is equivalent to the local GPU power density when power is dissipated internally and fully within the chip volume, with negligible external loss.

A representative 3D AI chip architecture with stacked GPU-HBM modules was modeled in COMSOL Multiphysics, as illustrated in Figure 1(a). The top thermal sink layer is used for heat dissipation. The GPU die layer is assumed to have a power density of 100 W/cm². The middle HBM stack contains a 5-layer configuration. The layer between the HBM and GPU is the interposer layer (Si or h-BN), the supporting substrate, and Through Silicon Vials (TSVs), respectively. TSVs are used not only to integrate the GPU to the memory and interposer, but also to provide the electrical and thermal connection between them. Figure 1(b) and (c) show the temperature distribution for the GPU-HBM architecture with different interposer layer materials. The top surface of the interposer is subjected to forced convection heating with an ambient heat transfer coefficient (h_amb) in a range of 150-350 W/(m²·K). The bottom of the chip substrate is cooled by natural convection with $h_b$ = 10 W/(m²·K). The boundary condition is applied based on Newton's Law of Cooling as follows[24]:

$$-k_s \frac{\partial T}{\partial n} = h(T - T_e) \qquad (2)$$

Where $T$ is the temperature at the solid surface [K]; $T_e$ is the external or ambient temperature [K]; $h$ is the heat transfer coefficient [W/m²·K]; $k_s$ is the thermal conductivity of the solid layers in AI chips [W/m·K]; $\frac{\partial T}{\partial n}$ is the temperature gradient normal to the surface [K/m].

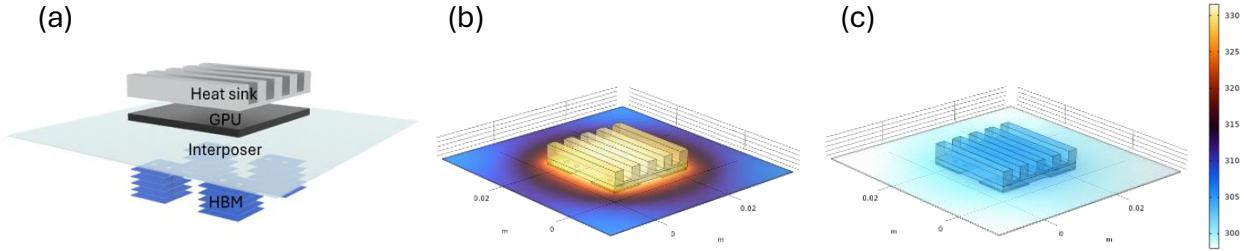

Figure 1. (a) Schematic diagram of the 3D AI chips integrating GPU and HBM modules, (b) COMSOL simulation result of junction temperature in the 3D AI chip with (b)Si interposer, (c)h-BN interposer.

## 3. Effects of HBM distribution on hotspot temperature

As illustrated in Figure 1(c), the GPU layer in the chip with an h-BN interposer exhibits a significantly lower junction temperature compared to the silicon-based counterpart. This confirms the superior heat-spreading capability of h-BN under high heat flux conditions. Building upon this observation, we further investigated how the spatial distribution of HBM modules influences the overall thermal behavior of the 3D AI chip. In 3D AI chips, the hotspot temperature is usually located in the GPU layer. The effects of varying the spatial distribution of 20 HBMs in the 3D AI chips were examined. The results are shown in Figure 2. For the GPU system 20 HBMs in one layer, a large hot-spot area is observed, with a temperature as high as 315 ºC. When 10HBMs/layer and two layers are designed, the hot-spot area is significantly reduced, and the hot-spot temperature is also slightly decreased. Interestingly, the hot spot temperature dropped more than 10 ºC when the layer-out is designed as 1 HBMs/layer × 20 layers, indicating optimal layout is critical for the 3D AI chip design.

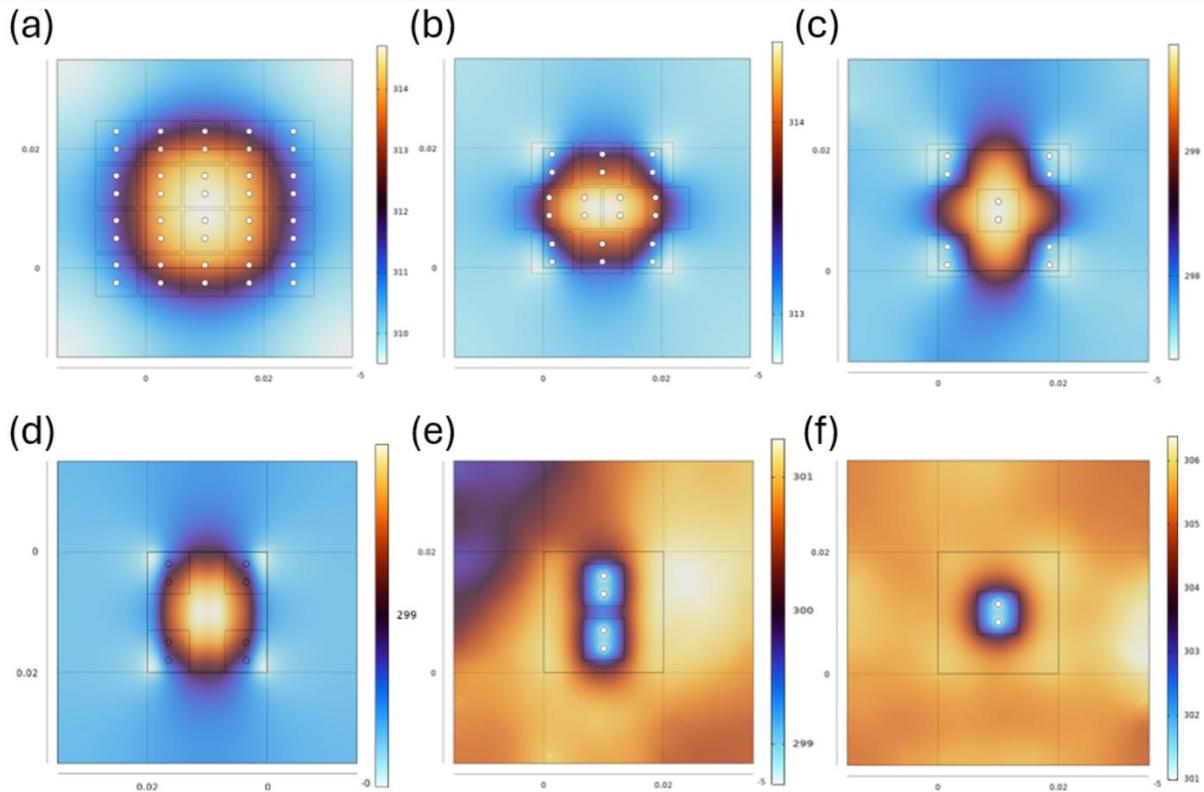

Figure 2. COMSOL simulation results of hotspot temperature in the GPU layer for varying spatial distribution of 20 HBMs connected with GPU by the TSVs. (a) 20 HBMs/layer ×1 vertical layer; (b) 10 HBMs/layer × 2 vertical layers; (c) 5 HBMs/layer × 4 vertical layers; (d) 4 HBMs/layer × 5 vertical layers; (e) 2 HBMs/layer × 10 vertical layers; (f) 1 HBMs/layer × 20 vertical layers.

As shown in Figure 3(a), when switching the layout design from 20 HBMs/layer with 1-layer interposer design to 10 HBMs/layer with 2-layer interposers, only a very slight decrease in hot-spot temperature is observed. The significant drop occurred for 4-layer designs with 5 HBMs per layer or a low number of HBMs per layer. The 5-layer design also showed similar performance, but there was only a very modest change in the hot-spot temperature for further spreading out the HBMs. That means optimal design can be achieved with 5HBMs/layer × 4 layers for cost effectiveness and miniaturization. This trend is attributed to the improved heat spreading and more uniform power dissipation enabled by multi-layer HBM integration. Among the configurations studied, distributing HBMs across four layers provides a favorable balance between thermal performance and design complexity.

Beyond the effects of HBM distribution, we further investigated how the h-BN interposer thickness impacts the thermal performance of the GPU, as illustrated in Figure 3(b). The simulation was built upon the optimal layer configuration(5 HBMs/layer × 4 layers) identified in Figure 3(a). As the thickness of the h-BN layer increases from 50 μm, a noticeable reduction in GPU surface temperature is observed. However, the rate of temperature reduction begins to level off at ~300 μm, indicating a saturation effect where further increments in h-BN thickness yield only marginal thermal improvements. Therefore, considering both thermal performance and material cost, a h-BN interposer thickness of approximately 300 μm is sufficient to achieve effective thermal management.

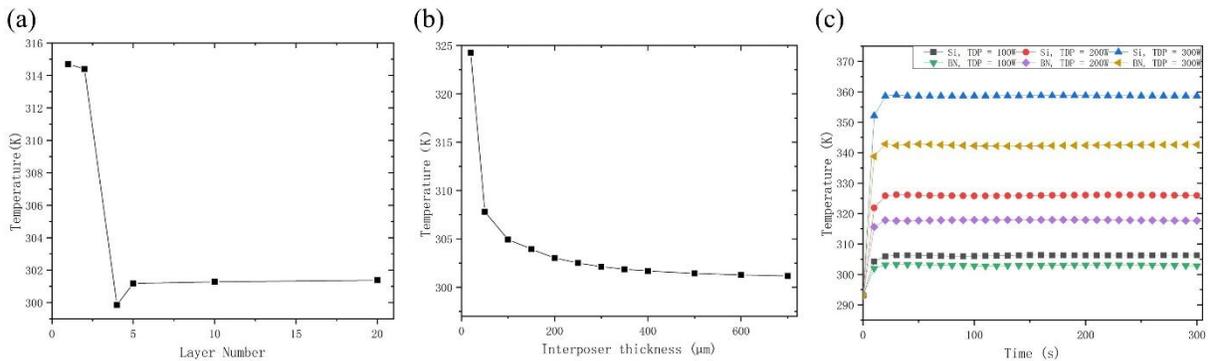

Figure 3. (a) Hotspot temperature in GPU as a function of the HBM layer number for varying spatial distribution of 20 HBMs, (b) surface temperature of GPU as a function of the h-BN interposer layer thickness with the structure of 5 HBMs/layer × 4 layers, and (c) evolution of GPU temperature as a function of time for different interposer materials under varying TDP values when interposer thickness = 300 μm with the structure of 5 HBMs/layer × 4 layers.

## 4. Effects of local GPU power density

This optimized configuration with 5 HBMs/layer × 4 layers and a 300 μm thick h-BN interposer was then used in the subsequent COMSOL simulation. The performance of h-BN and Si as interposer materials under different GPU thermal design power (TDP) values of 100 W, 200 W, and 300 W were evaluated, respectively.

The effect of local power density on GPU temperature follows approximately[23, 25]:

$$T_{GPU} \propto \frac{\dot{q}_g \cdot L^2}{k_{eff}} \qquad (3)$$

Where $\dot{q}_g$ is the local power density [W/m³]; $L$ is the characteristic length of the thermal path, such as chip thickness [m]; and $k_{eff}$ is the effective thermal conductivity of the stack [W/m·K].

The GPU temperature follows a transient heating curve due to the nature of Joule heating before reaching a steady state. Briefly, in the first 10 seconds, GPU temperature is positively related to the local power density and rises steeply. The rise rate is dependent on the local power density, heat capacity, and initial thermal resistance. After 10 seconds, the hotspot temperature reaches an equilibrium, when a thermal equilibrium between the power input and heat output is achieved. Compared to a traditional silicon interposer, the designed h-BN interposer reduces hotspot temperatures by 20 °C, equivalent to approximately a 6% reduction in thermal resistance under a heat flux of 300 W/cm². The h-BN interposer achieves better heat dissipation than the Si interposer across all tested TDP values. This improvement not only enhances device reliability and performance stability but also extends operational lifetime under high-power 3D AI chip workloads. For example, such an improvement could cut down Complementary Metal Oxide Semiconductor (CMOS) power leakage by 22%.

## 5. Conclusion

This study demonstrates the thermal advantages of employing h-BN as an interposer material in 3D GPU-HBM architectures. Compared to conventional Si interposers, h-BN interposers significantly reduce hotspot temperatures and improve temperature uniformity due to their superior in-plane thermal conductivity and electrical insulation. COMSOL simulations reveal that distributing HBM modules across multiple layers and optimizing the thickness of the h-BN interposer can effectively mitigate thermal stress and enhance overall heat dissipation. These findings provide practical insights for the early-stage thermal design of next-generation AI chips operating under high power densities.

## 6. Acknowledgement

The authors are grateful to the facility and funding support from Texas A&M University.

*Conflict of interest statement:* On behalf of all authors, the corresponding author states that there is no conflict of interest.

*Data availability statement:* Data will be made available on reasonable request.

*Author contribution:*

Eric Han Wang: conceptualization, validation, writing – original draft.

Weijia Yan: software, methodology, visualization, writing – review and editing, supervision.

Ruihong Huang: resources, writing – review and editing, supervision.